# Systematic Review of Methods and Prognostic Value of Mitotic Activity – Part 1: Feline Tumors

Christof A. Bertram [1], Taryn A. Donovan [2], Alexander Bartel [3]

[1] University of Veterinary Medicine Vienna, Vienna, Austria

[2] The Schwarzman Animal Medical Center, New York, USA

[3] Freie Universität Berlin, Berlin, Germany

Corresponding author:

Alexander Bartel, Department of Veterinary Medicine, Institute for Veterinary Epidemiology and Biostatistics, Freie Universität Berlin, Koenigsweg 67, Berlin, 14163 Berlin, Germany. Email: Alexander.Bartel@fu-berlin.de

## Abstract

Increased proliferation is a key driver of tumorigenesis, and quantification of mitotic activity is a standard task for prognostication. The goal of this systematic review is scholarly analysis of all available references on mitotic activity in feline tumors, and to provide an overview of the measuring methods and prognostic value. A systematic literature search in PubMed and Scopus and a manual search in Google Scholar was conducted. All articles on feline tumors that correlated mitotic activity with patient outcome were identified. Data analysis revealed that of the eligible 42 articles, the mitotic count (MC, mitotic figures per tumor area) was evaluated in 39 instances and the mitotic index (MI, mitotic figures per tumor cells) in three instances. The risk of bias was considered high for most studies (26/42, 62%) based on small study populations, insufficient details of the MC/MI methods, and lack of statistical measures for diagnostic accuracy or effect on outcome. The MC/MI methods varied markedly between studies. A significant association of the MC with survival was determined in 21/29 (72%) studies, while one study found an inverse effect. There were three tumor types with at least four studies and a prognostic association was found in 5/6 studies on mast cell tumors, 5/5 on mammary tumors and 3/4 on soft tissue sarcomas. The MI was shown to correlate with survival by two research groups, however a comparison to the MC was not conducted. An updated systematic review will be needed with of new literature for different tumor types.

# Introduction

Increased tumor cell proliferation through self-sufficiency in growth and avoidance of growth-inhibitory signals is a key driver of tumorigenesis (hallmark of cancer). Mitotic activity is a highly relevant measure for the growth fraction of tumor proliferation [55] and is thus generally expected to correlate with more aggressive biological behavior and less favorable patient outcome for malignant tumor types. Quantification of mitotic figures (cells undergoing cell division visible in histological sections) is a standard task for tumor prognostication in veterinary pathology due to its high practicability and assumed high prognostic value.[16,32] However, given the extensive availability of feline oncologic literature, identification of the recommended methods for measuring mitotic activity as well as the prognostic relevance of these tumor parameters can be difficult to ascertain for each individual tumor type.

There are two broad categories of measurement methods for mitotic activity, namely the mitotic count (MC) and the mitotic index (MI). While the MC represents the number of mitotic figures per tumor area, the proportion of mitotic figures among all tumor cells is measured by the MI.[32,33] Descriptions of the measuring methods of mitotic activity in oncologic research must be adequately detailed such that others can reliably and accurately reproduce the methods and data can be compared. Recent guidelines have defined key aspects of the MC including the region of interest (ROI) within the tumor section, the size of the ROI in mm², the spatial arrangement of the fields of view (high power field, HPF) within the ROI, and identification criteria of mitotic figures.[16,32] Standardized methods for the MI have not been proposed for veterinary oncology. A summary of the methods applied in previous studies is needed as tumor-type specific practice may differ from standardized methods.

Regardless of the critical biological role of tumor cell proliferation in cancer development, the prognostic utility of mitotic activity in some feline tumors was not justified.[28,49,61] Potential explanations include the true lack of a prognostic relevance for mitotic activity in a particular tumor type and/or the use of non-standardized, inconsistent or inaccurate study methods or non-representative study populations. Replication studies are needed due to the high risk of bias of observational studies, and to validate the results in different study populations. Only then final conclusions on the relevance of prognostic tests can be drawn. Currently, no evidence-based recommendations exist to confirm or deny whether mitotic activity should be evaluated in the individual feline tumor types as a routine prognostic test.

This systematic review intends to satisfy the need for a scholarly overview of the methods and prognostic value of measuring mitotic activity in feline tumors. An extensive literature search was conducted to find all studies that correlate the MC or MI with any type of patient outcome.

# Material and Methods

## Literature search

A literature search protocol was developed based on the PRISMA statement.[37] The literature search consisted of 1) a systematic search in two databases (PubMed and Scopus) using predefined search terms and 2) a manual search in a database (Google Scholar) to ensure literature saturation (Fig. 1). All identified references were screened for eligibility for study inclusion by two authors (CAB and TAB).

For the systematic literature search one author searched Pubmed (1950 to present) and Scopus (1970 to present) on June 9th 2022. The search string was build based on two topics for Who (animal) and What (prognostic test) resulting in the following search strings: (cat OR cats OR feline) AND (mitotic count OR mitotic index). The records identified in each database using these six combinations were exported to Endnote (Version X9.9.9) and sorted alphabetically based on their title in order to easily identify duplicates, which were subsequently removed. Two separate eligibility screening steps, the title-abstract and full-text screening, were done in the web application Rayyan [36] by the two literature reviewers in a blinded manner with the exception of one article that was written in German, which was only reviewed by CAB. Any disagreement between the two literature reviewers in the two screening steps were solved by joint full text review and discussion. The inclusion / exclusion criteria for the two screening steps are summarized in Table 1. Only papers that reported statistical results on the prognostic value of mitotic activity as a solitary parameter (i.e. not in combination with other parameters, such as in a grading system) or provided individual patient data were included in this systematic review. Indirect assumptions on the prognostic value by correlation of the mitotic activity with other established prognostic parameters are not reliable and were not considered eligible.

Manual literature search was conducted from May to June 9th 2022. The database Google Scholar was searched with free search terms (including terms such as "outcome", "prognosis" and relevant tumor types). Additionally, the citing references of the articles of interest ("cited by" search in Google Scholar) were examined. Articles of potential interest were identified through the title and abstract, and the full text was screened for the same eligibility criteria as stated above. The records and full text of those papers that met the eligibility criteria were extracted and duplicates from the systematic literature search were removed. Eligibility for inclusion of the remaining articles was verified by a second literature reviewer.

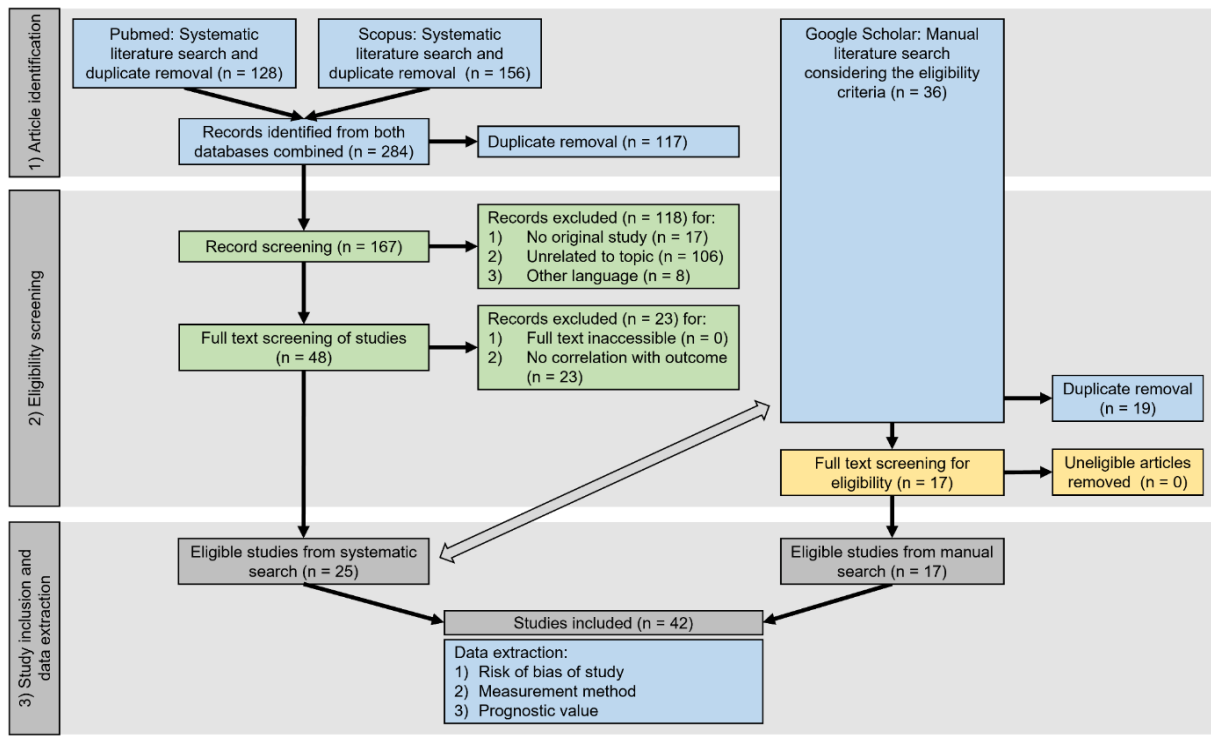

**Figure 1.** Flow diagram of systematic (on the left) and manual (on the right) literature search, eligibility screening, and study inclusion and data extraction. The tasks in the blue boxes were conducted by the primary literature reviewer (CAB), the tasks in the green boxes were conducted by two literature reviewers in a blinded manner, and the tasks in the yellow boxes were conducted by a second literature reviewer (TAB).

**Table 1.** Summary of the eligibility criteria for the two screening steps of the systematic literature search.

| Screening Step | Decision category | | Inclusion criteria | Exclusion criteria |
|---|---|---|---|---|
| Title-abstract | 1) Study design | | Original study, peer-reviewed | Case-reports, reviews |
| | 2) Topic | a) Species | Cat / feline | Other species |
| | | b) Tumor | Spontaneous tumors | Experimentally induced tumors |
| | | | Malignant tumors with potential for metastasis | Benign tumors |
| | | c) Prognostic test | Mitotic count (MC), mitotic index (MI) | No mitotic activity measurement |
| | 3) Language of main text | | English or German | Other language |
| Full text | 1) Article accessibility | | Article accessible | Article inaccessible |
| | 2) Topic | a) Patient outcome | Correlation of the MC/MI with survival, tumor progression, metastasis, or recurrence | No correlation of the MC/MI with patient follow-up |

### Data extraction and risk of bias (RoB)

Data extraction of the relevant information from the articles that were included in the systematic review was performed by CAB. For each study the citation information, year of publication, and journal type (journal focusing on animal pathology, i.e. Vet Pathol, J Comp Pathol and J Vet Diagn Invest, versus other journals), and tumor type evaluated were recorded. Subsequently, information regarding the risk of bias (RoB), mitotic activity measurement method and prognostic value (association of the MC/MI with patient follow-up) was extracted. Studies on the mitotic count and mitotic index were analyzed separately.

RoB from each included article was evaluated regarding the information on mitotic activity. New decision criteria were developed based on previous recommendations.[6,26,32,51,57] These new criteria were intended to be straightforward, applicable for this specific systematic review, and concede to the current practice of prognostic studies. We grouped the decision criteria into four domains that are critical when conducting a prognostic study: 1) study population, 2) outcome assessment, 3) mitotic activity measurement method, and 4) data analysis. The overall RoB was rated by combining all four domains. While high risk of domain 3 was considered less severe compared to the other domains (although being highly important for replicability of the results) the combined score could only be one level higher than the worst score.

## Results

### Study selection

The literature selection process is summarized in Figure 1. From 167 unique references found by the systematic database search 25 articles remained eligible. Disagreement between the two literature reviewers only occurred for one the articles during the title-abstract screening steps. The manual literature search identified 17 additional articles. Finally, 42 articles were included in this systematic review.[2,7,8,10-15,18,19,21-24,27-30,34,35,38-50,52,54,56,58-62]

### Study characterizations

All articles were written in English except for one article, written in German.[24] Sixteen of the 42 (38%) articles were published in journals with a focus on veterinary pathology (Vet Pathol, N = 13; J Comp Pathol, N = 1; J Vet Diagn Invest, N = 1) and 26/42 articles (62%) in other journals.

The mitotic activity measurement methods were described in 40/42 included papers (95%) and the measuring method was not specified in two papers [28,29] (5%). According to our definition, the MC, i.e. the number of mitotic figures per tumor area, was described in 37 papers. We assume that the two papers without specified methods also conducted the MC as the data was taken retrospectively from medical records (total: 39/42 papers; 93%). Three [40,50,54] (7%) had performed the MI, i.e. the percentage of mitotic figures per number of tumor cells. The number of publications on mitotic activity increased notably over the last decade (Fig. 2).

Whereas the three studies on the MI used the correct terminology in their paper, the 39 studies on the MC used various and sometimes multiple terms including MC (N = 16, 41%), MI (N = 16, 41%), number of mitosis (N = 4, 10%), mitotic rate (N = 3, 8%), mitosis (N = 2, 5%), and mitotic activity (n = 1, 3%). In January 2016, a guest editorial in Vet Pathol [33] pointed out the proper terminology for MC and MI. While only 13% (3/23) of the articles published before 2017 used the correct term "MC", 81% (13/16) published after 2016 used the correct terms. Incorrect use of terminology after 2017 occurred only in non-pathology-focus journals.

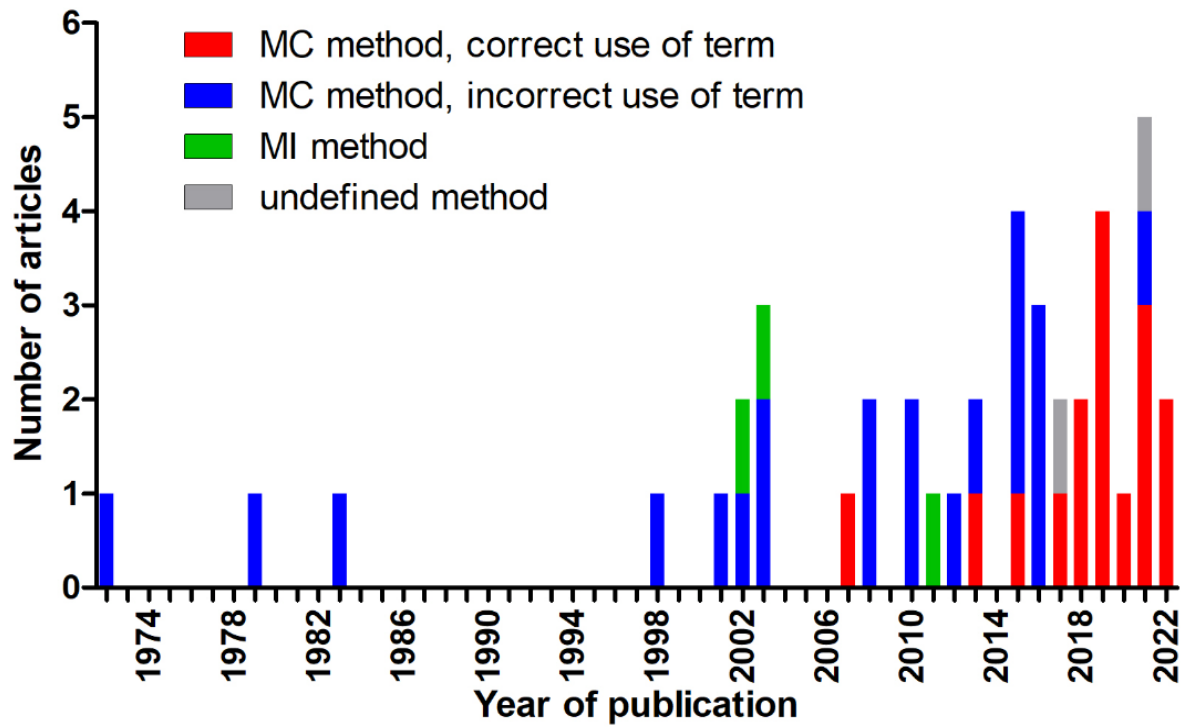

**Figure 2.** Stacked bar chart of the number of publications on the mitotic count (MC) and mitotic index (MI) included in this systematic review separated by year of publication. Any term other than MC was considered incorrect for measuring the number of mitotic figures per unit area. The year 2022 only includes January to June 9th.

### Descriptive summary

#### Mitotic count (MC)

The 39 articles on the MC investigated a variety of different tumor types or tumor groups. Tumor types/groups with multiple studies included mast cell tumors (N = 9; cutaneous, N = 7, pleomorphic cutaneous, N = 1; intestinal, N = 1), soft tissue sarcoma (N = 6; injection-site sarcoma, N = 3; cutaneous, N = 1; piloleiomyosarcoma, N = 1; fibrosarcoma of mostly skin, and few cases of oral cavity and bone, N = 1), malignant mammary tumors (n = 5), and melanocytic tumors (N = 5; non-ocular, N = 3; nasal planum, N = 1; iris, N = 1).

**Risk of bias (RoB) of the studies on the MC**

The RoB for all studies is summarized in Table 2. The bias of the study population was mostly influenced by small case numbers with outcome, which was particularly relevant for tumor types with a single study (low: 0, moderate: 6, high: 10) as compared to tumor types with multiple studies (low: 7, moderate: 11, high: 5). For outcome assessment, less stringent criteria were applied resulting in a lower number of high risk studies. This is mostly due to the decision to assign a moderate RoB to the procedure of following up by questionnaires or medical records, which was by far the most common approach. Only one study conducted a postmortem examination after death of the patient [7] and two studies conducted prospective clinical follow-up or questionnaires at regular intervals.[7,61] MC methods were often incompletely described, as detailed in the next section, leading to a high RoB in 24/39 articles (62%). Data analysis (correlation of the MC with outcome) was often restricted to the p-value approach (mostly log-rank test), which does not allow evaluation of the discriminant ability of the test.[6] Of 29 studies that statistically correlated the MC with survival, 11 (38%) provided only results of tests of significance (p-value approach) and results of other statistical tests or graphical illustrations were not available. Of these 11 articles, 4 (36%) reported the p-value and 7 (64%) only stated that the p-values were above the level of significance (without providing the actual value). Other statistical tests such as median survival time, Kaplan-Meier curves, hazard ratios or odds ratios, ROC curves and AUC, sensitivity and specificity were inconsistently or rarely conducted.

**Table 2.** Summary of the risk of bias (RoB) for each evaluated domain (D1-4) and overall RoB for all studies on the mitotic count (MC) in feline tumors combined (N = 39).

| RoB | Number and percent of articles | | | | |
| --- | --- | --- | --- | --- | --- |
|  | D1: Study population | D2: Outcome assessment | D3: MC method | D4: Data analysis | Overall (D1-4) |
| ⊕ Low | 7 (18%) | 3 (8%) | 6 (15%) | 2 (5%) | 1 (3%) |
| ◯ Moderate | 17 (44%) | 28 (72%) | 9 (23%) | 15 (38%) | 12 (31%) |
| ⊖ High | 15 (38%) | 8 (21%) | 24 (62%) | 22 (56%) | 26 (67%) |

**MC methods**

The MC values were taken from medical records (pathology reports) in 3/39 (8%) studies.[8,28,29] The remaining 36/39 (92%) studies determined the MC values based on their study protocol using HE-stain or, in one instance, HE-Saffron-stain.[12] Bleaching of melanin pigment was performed for melanocytic tumors if required.[38,41,48] In five studies,[34,38,39,48,56] histopathological evaluation was done by multiple (2 – 4) pathologists with consensus in disagreed cases (N = 2), simultaneous evaluation (N = 1) or calculation of the mean value (N = 1; unknown for one study). None of the studies reported the use of digital microscopy or automated image analysis.

The specific MC methods are summarized in Fig. 3. Generally, a somewhat higher proportion of articles published after 2017 and in pathology-focused journals described the individual aspects of the MC methods.

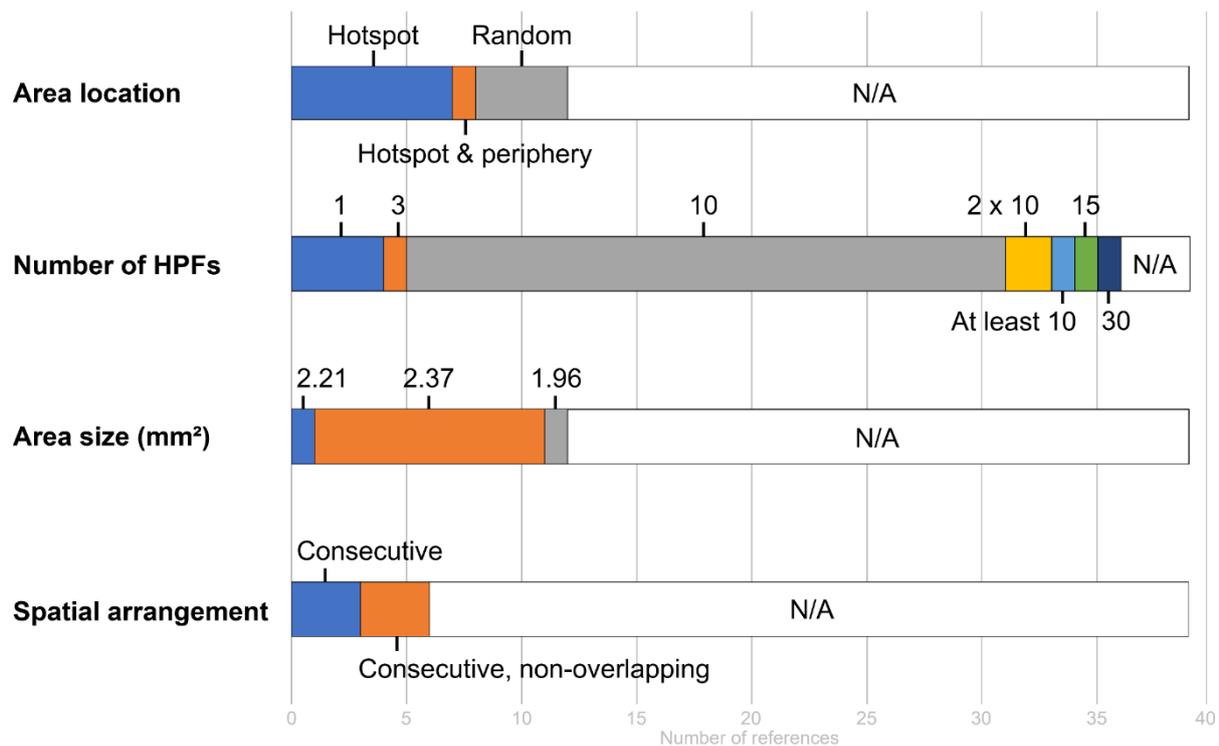

**Figure 3.** Stacked bar chart of the mitotic count methods used in 39 eligible studies on feline tumors regarding the area location of the enumerated field(s) within the tumor section, number of high-power fields (HPFs) enumerated, area size in mm² (based on the field number of the light microscope), spatial arrangement of the individual fields of view.

N/A, not available

**Prognostic value of the MC**

The 39 references included outcome information by providing statistical results in the paper (N = 29), tables with individual patient data (N = 6), or both (N = 4). Information on survival time was available in 32 articles, tumor progression (metastasis and/or recurrence) in 8 articles, metastasis in 6 article, and recurrence in 8 articles. The evaluated number of cases with follow-up per study ranged between 4 and 342 (mean: 44; median: 30).

Prognostic cut-off values were provided in 23 studies (59%), whereas only 12 (52%) indicated how this classification was created: median of the MC values (N = 4 [34,46,47,61]), mean of the MC values (N = 1 [43]), tertiles of the MC values (N = 1 [24]), receiver operating characteristic (ROC) curve (N = 4 [12,14,39,44]), based on a previous literature (N = 1 [18]), or "based on data analysis and literature" (N = 1 [48]).

Regarding survival, 20/28 (71%) found a significant association (p-value approach) of higher mitotic counts being associated with shorter survival and 8/28 (29%) did not (Table 3; four additional studies only provided individual patient data). Hammer et al.[21] experienced an inverse effect in salivary gland tumors with higher MC values being significantly associated with longer survival. There are 4 tumor types with at least 3 studies evaluating survival. Higher MCs were associated with a shorter survival time or lower survival rates in 5/6 studies on mast cell tumors of the skin, 5/5 studies on mammary tumors, 3/4 studies on (sub)cutaneous soft tissue sarcoma, and 2/3 studies on non-ocular melanocytic tumors (Fig. 4). The area under the ROC curve (AUC) was only reported for two studies on mast cell tumors suggesting a high discriminant ability (AUC = 0.79 and 0.92).[14,44]

Tumor progression was significantly correlated with higher MCs in 4/6 studies (67%; Table 3). A prognostic significance of the MC regarding occurrence of metastasis was found in 2/3 studies (67%, Table 3). Tumor recurrence was significantly associated with higher MCs in 4/8 studies (50%; Table 3). While recurrence was often evaluated for soft tissue sarcoma (N = 5), only 2/5 (40%) found a prognostic significance for the MC.

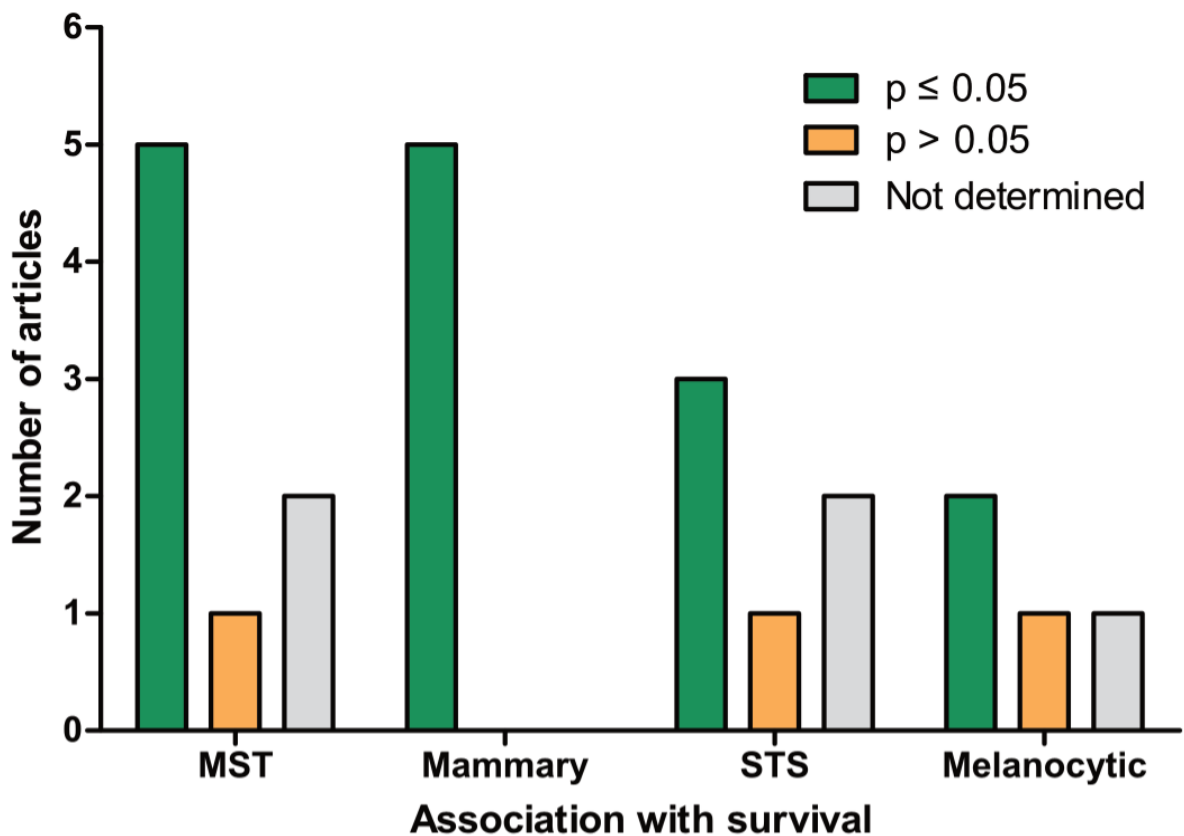

**Figure 4.** Prognostic significance for survival (p-value approach) for the feline tumor types with more than three articles. Study results with a p-value of ≤ 0.05 are considered significant.

MST, cutaneous mast cell tumors; mammary, malignant mammary tumors, STS, (mostly) cutaneous soft tissue sarcoma; melanocytic, non-ocular melanocytic tumors;

**Supplemental Table S6.** Summary of each evaluated study regarding the prognostic significance of the mitotic count in different feline tumor types based on the p-value approach or conclusion by the authors.

| Tumor type | Reference (author and year of publication) | Association suggested with | | | |
|---|---|---|---|---|---|
| | | Survival | Progression | Metastasis | Recurrence |
| Ceruminous gland adenocarcinoma | Bacon et al., 2003 | Yes | – | – | – |
| Hemangiosarcoma | Johannes et al., 2007 | Yes | – | – | – |
| Lymphoma | Santagostino et al., 2015 | No | – | – | – |
| Mammary tumor | Dagher et al., 2019 | Yes | – | – | – |
| | Mills et al., 2015 | Yes | – | – | – |
| | Rosen et al., 2020 | Yes | Yes | – | – |
| | Weijer and Hart, 1983 | Yes | – | – | No |
| | Weijer et al., 1972 | Yes | – | – | – |
| Mast cell tumor, cutaneous | Dobromylskyj et al., 2015 | Yes | – | – | – |
| | Johnson et al., 2002 | – | – | – | Yes |
| | Lepri et al., 2003 | – | Yes | – | – |
| | Melville et al., 2015 | Yes | – | – | – |
| | Molander-McCrary et al., 1998 | No | – | – | – |
| | Sabattini and Bettini, 2019 | Yes | – | – | – |
| | Sabattini and Bettini, 2010 | Yes | – | – | – |
| | Sabattini et al., 2013 | Yes | Yes | – | – |
| Mast cell tumor, intestinal | Sabattini et al., 2016 | Yes | – | – | – |
| Malignant maxillary tumors | Liptak et al., 2021 | Yes | Yes | – | – |
| Melanocytic tumors, non-ocular | Chamel et al., 2017 | No | – | No | – |
| | Pittaway et al., 2019 | Yes | – | – | – |

| Tumor type | Reference (author and year of publication) | Association suggested with | | | |
|---|---|---|---|---|---|
| | | **Survival** | **Progression** | **Metastasis** | **Recurrence** |
| | Sabattini et al., 2018 | Yes | – | – | – |
| | Reck and Kessler, 2021 | IPD | IPD | – | – |
| Melanoma, iris | Wiggans et al., 2016 | – | – | Yes | – |
| Merkel cell tumor | Sumi et al., 2018 | IPD | IPD | IPD | – |
| Oral tumors | Saverino et al., 2021 | IPD | – | IPD | – |
| Osteosarcoma | Dimopoulou et al., 2008 | – | – | – | Yes |
| Pancreatic tumor | Linderman et al., 2013 | No | No | – | – |
| Plasma cell tumor | Craig and Lieske, 2022 | IPD | – | IPD | – |
| Progressive histiocytosis | Coste et al., 2019 | No | – | – | – |
| Salivary gland tumor | Hammer et al., 2001 | Inverse | – | – | – |
| Soft tissue sarcoma, (mostly) skin | Bostock and Dye, 1979 | Yes | – | – | No |
| | Guidice et al., 2010 | – | – | – | No |
| | Kamenica et al., 2008, | No | – | – | No |
| | Porcellato et al., 2017 | Yes | – | – | Yes |
| | Guisado and Castro, 2022 | – | – | Yes | Yes |
| | Dobromylskyj et al., 2021 | Yes | – | – | – |
| Squamous cell carcinoma, oral | Yoshikawa et al., 2016 | No | No | – | – |
| | Yoshikawa et al., 2012 | No | – | – | – |
| Squamous cell carcinoma, cutaneous | Rodriguez Guisado et al., 2021 | Yes | – | – | – |

## Mitotic index (MI): risk of bias (RoB), methods, prognostic value

Three papers on the MI examined feline mammary carcinoma.[40,50,54] The overall RoB was judged to be moderate for all three studies.

Two studies were conducted by the same research group who used the same MI methods and presumably the same study population.[40,50] These two studies used toluidine blue stain for enhancement of mitotic figures (as suggested by a previous study [9]). For the MI (mitotic figures per 100 tumor cells), 10 HPFs at 250x magnification were selected in a hotspot location and presumably photomicrographs were taken. Utilizing an image cytometry software, the total nuclear area and mean nuclear area of tumor cells was determined, the quotient of which was used to estimate the number of tumor cells.

The third study determined the MI (mitotic figures per 1000 tumor cells) from 1000 tumor cells in 8-10 HPFs selected at the periphery and hotspot tumor location.[54] The MI was probably determined in Ki67-immunolabeled sections.

The three studies determined a significantly shorter tumor-specific survival time for cases with higher MI. As the prognostic cut-off value, all studies used the median MI values of their study populations.

## Discussion

This systematic review analyzes measurement methods and prognostic value of the MC and MI in feline tumors. The evaluated studies found a significant association of poor outcome with higher MCs in 50% - 71% (depending on the outcome metric) instances and with higher MIs in all three instances. Overall, this suggesting a prognostic relevance of mitotic activity in feline tumors. However, the ability to derive conclusions was limited. A general finding of the systematic review is that there is a paucity of literature for most tumor types, there are various mitotic activity measurement methods, and many studies lack relevant information to interpret the prognostic value of mitotic activity. As observational studies have bias in the study populations (case selection and outcome assessment) and as there is high rater variability in enumerating mitotic figures,[3,4,16] multiple studies are crucial to confirm, modify or reject research findings.[31,53] There is a marked increase in the number of articles on the prognostic relevance of the MC over the last decade and our tumor type-specific conclusions may be updated in a couple of years when more validation studies are available. Adherence of future prognostic studies to recommendations for conducting prognostic studies and standardized MC methods will facilitate systematic reviews and allow for meta-analysis.

Not surprisingly, most studies evaluated the MC, which is much more practical than the MI. However, the terminology for this measurement method was inconsistently used in previous studies. Terms such as MI, mitotic rate or mitosis should be avoided to denote enumeration of mitotic figures in a certain tumor area. According to recent definitions, the MI indicates that the mitotic density was determined relative to the tumor cell density.[33] "Rate" is defined as the number of events over time (such as the heart rate: beats per 1 minute). Mitosis is the process of cell division, which cannot be seen under the light microscope. The morphology of dividing cells within the M phase of the cell cycle are visible to pathologists as structures called "mitotic figures".

The mitotic count methods were quite variable and relevant information was often lacking. Since 2016, some commentaries and guidelines have been published [16,32,33] to increase awareness of the veterinary community to critical methodological aspects of the MC needed for standardization. Particularly, the appropriate measure for the area size evaluated (mm² instead of HPFs) was highlighted. Of note, recent articles provide information concerning area size more often, showing that researchers are increasingly becoming aware of the variability of the MC methods. A recent guideline developed by the VCGP group,[32] provides standard recommendations for each of the critical methodological aspects of the MC (hotspot location; consecutive, non-overlapping fields of view; 2.37 mm² area size), which were the most common methods used in the feline studies with provided information. However, other methods were not uncommon, and it is still largely unknown which method is the best approach from a prognostic standpoint. While the best method is still debatable, we highlight that studies need to report the precise methods on all aspects of the mitotic count: location within tumor, spatial arrangement of the field of view, and area size in mm².

A prognostic value of the MC for survival could only be evaluated for a few tumor types/groups with multiple studies. Using the p-value approach, a significant association with survival was found by all or most studies on cutaneous mast cell tumors, mammary tumors and soft tissue sarcoma. For all the other tumor types/groups, the prognostic value of the MC remains unproven considering the lack of validation. A difficulty regard the comparison/combintion of studies is that many studies evaluated a heterogeneous group of tumors (such as fibrosarcoma of skin, bone and the oral cavity [7]), while other studies evaluated specific tumor entities (such as cutaneous piloleiomyosarcoma [19]). Combined evaluation of studies on the different tumor subtypes and/or locations was necessary in the present study due to the paucity of studies.

Interpretation of the prognostic value based on the p-value approach has considerable limitations as it depends upon the case numbers of the study population and does not allow evaluation of the discriminant ability of the MC.[6] Considering the case numbers relative to the p-values, the discriminant ability of mammary tumors might be somewhat smaller than for other tumor types. Another limitation regarding the p-value approach was that many studies did not provide the actual values when they were above the level of significance, which markedly hindered the ability to interpret the results as the p-values above the level of significance can range between 0.05 and 1.0. Statistical analysis that better described the discriminant ability (such as hazard ratios, AUC, sensitivity and specificity)[6] graphical illustrations (such as the Kaplan-Meier curve, scatterplots or ROC curves) were inconsistently reported, which precludes in depth evaluation of the prognostic relevance of the MC.

The MI was reported only in three studies on feline mammary tumors,[40,50,54] two of which presumably used the same study population.[40,50] Whereas these two research groups suggest a prognostic value for mammary tumors, further studies are needed that provide statistical tests for discriminant ability and evaluate further tumor types. Mammary tumors seem to be a good candidate for the MI due to the variable tumor cell density resulting from cystic spaces, sclerosis, and necrotic areas. Adjustment by the tumor cell density may reflect the mitotic activity of the tumor better and may provide a more accurate prognosis, however a prognostic improvement as compared to the routine MC has not been shown for feline mammary tumors. An alternative solution to adjust for variable tumor cell density is the volume-corrected mitotic counts,[20] which has, however, not been evaluated as a prognostic test for animal tumors thus far.

The limitation of the MI is the additional time investment for counting the number of tumor cells, which hampers application in a routine diagnostic workflow. We consider automated image analysis very promising to facilitate the MI in the future, when it has become widely available. An increasing number of laboratories use digital microscopy for their diagnostic workflow,[5] and development of deep learning-based image analysis algorithms, including those that can segment and count tumor cell nuclei,[17,25] are of high research interest. Automated enumeration of tumor cell nuclei, possibly in combination with algorithmic detection of mitotic figures,[1,3] would allow a routine application of the MI.

## Conclusion

Mitotic activity is often considered one of the most useful histological prognostic tests for tumors. In cats, however, there is a paucity of literature, considerable RoB, and high variability of the measurement method in the current literature for both the MC and MI. More than two-thirds of the studies found a prognostic significance of the MC for patient survival, indicating a general relevance of this prognostic test. However, sufficient evidence for the prognostic value only exists for few tumor types, namely cutaneous mast cell tumors, mammary tumors, and possibly for soft tissue sarcoma. For other tumor types, validation studies are lacking. The MI has so far only been evaluated only for feline mammary carcinoma, while a prognostic comparison to the MC is lacking.

## Acknowledgements
None

## Declaration of conflict of interest
None

## Funding
None

## Authors' Contributions

CAB, TAD and AB designed the experiment; CAB and TAD performed the systematic review; CAB performed data extraction from the articles; the manuscript was written by CAB with contributions by all authors.